\documentclass[aps,twocolumn,amsmath,amssymb,showpacs,showkeys,floatfix,nofootinbib]{revtex4}
\usepackage{graphicx}
\usepackage{amssymb} %%%%%%%%%%%%%%%%%%%%%%%%%%%%%%%%%%%%%%%%%%%%%%%%%%%%
\begin{document}

%%%%%%%%%%%%%%%%%%%%%%%%%%%%%%%%%%%%%%%%%%%%%%%%%%%%%%%%%%%%%%%%%%%%%%%%%%%

%\begin{frontmatter}

\title{Connecting the pygmy dipole resonance to the neutron skin}

\author{V. Baran, $^{1}$ M. Colonna,$^{2}$ M. Di Toro,$^{2,3}$ A. Croitoru,$^{1}$ and D. Dumitru$^{1}$}
\affiliation{$^{1}$ Faculty of Physics, University of Bucharest, Romania},
\email{baran@ifin.nipne.ro}
\affiliation{$^{2}$ Laboratori Nazionali del Sud, INFN, I-95123 Catania, Italy},
\affiliation{$^{3}$ Physics and Astronomy Department, University of Catania, Italy},
%%%%%%%%%%%%%%%%%%%%%%%%%%%%%%%%%%%%%%%%%%%%%%%%%%%%%%%%%%%%%%%%%%%

\begin{abstract}
We study the correlation between the neutron skin development and the low-energy dipole response
associated with the pygmy dipole resonance (PDR) in connection with the properties of symmetry energy. 
We perform our investigation within a microscopic transport model based on the Landau-Vlasov kinetic equation
by employing three different equations of state in the isovector sector. 
Together with the giant dipole resonance (GDR) for all studied systems, 
we identify a PDR collective mode whose energy centroid is very well described 
by the parametrization $E_{PDR}=41 A^{-1/3}$.
A linear correlation between the energy weighted sum rule (EWSR) associated to PDR 
and the neutron skin thickness is evidenced. An increase of $15 MeVfm^2$ of
EWSR, in correspondence to a change of $0.1fm$ of the neutron skin size, is obtained. We conjecture that different nuclei 
having close neutron skin sizes will exhaust the same EWSR in the pygmy region.
This suggests that a precise experimental
estimate of the total EWSR exhausted by the PDR allows the determination of the neutron skin size,
constraining the slope parameter of the symmetry energy. 
\end{abstract}
%\prod
%\noindent


%\keywords{
%Pygmy Dipole Resonance, Giant Dipole Resonance, Landau theory of Fermi liquids, Symmetry energy
%}
%\end{quote}}

%\date{\today}
\maketitle

\section{Introduction}

The nuclear symmetry energy, which originates from both Pauli correlations and the specific features of 
nuclear forces, accounts for the effects related to
the difference between the number of protons $Z$ and neutrons $N$ of the system. 
It appears in the expression of total energy per particle, 
$\displaystyle \frac{E}{A}(\rho, I) =\frac{E}{A}(\rho) +\frac{E_{sym}}{A}(\rho) I^2$, factorizing 
the isospin parameter $ \displaystyle I=\frac{N-Z}{A}$, where $\rho$ is the nucleon density.
Several features of atomic nuclei \cite{barPR2005,baoPR2008} and neutron stars \cite{stePR2005} are determined by
this quantity and one of the major tasks of recent experimental and theoretical investigations 
is to determine a consistent density parametrization of the symmetry energy 
which can provide a unified picture of nuclear properties below saturation as well as at large compression  
of asymmetric nuclear matter \cite{ditJPG2010}.

The fragmentation facilities at  GANIL, GSI, MSU and RIKEN, allowing for the study of very neutron rich
systems, stimulated new investigations along this direction. In this context, understanding the exotic modes of excitation 
\cite{paaRPP2007} and the role of the
neutron skin on the collective dynamics in nuclei far from stability is a challenge in modern nuclear physics \cite{isaPRC1992,carPRC2010,wiePPNP2011}.
Indeed, several experiments performed during the past 10 years reported
the occurrence of an electric dipole ($E1$) response well below the giant dipole resonance (GDR), more
clearly evidenced in neutron rich nuclei \cite{harPRL2000,harPRC2002,adrPRL2005,wiePRL2009,kliPRC2007,tamPRL2011,konPRC2012},
see Refs. \cite{aumPS2013,savPPNP2013} for recent overviews. It manifests as 
a resonant-like shape exhausting few percentages of the dipolar energy wighted sum rule (EWSR) and its controversial
nature attracted a considerable interest for theory too \cite{paaJPG2010}. The pygmy dipole resonance (PDR) was interpreted as a collective 
motion in phenomenological, hydrodynamic descriptions \cite{mohPRC1971}, in 
nonrelativistic microscopic models \cite{coPRC2009,lanPRC2011,mazPRC2012,yukNPA2012} or in transport models
\cite{urbPRC2012,barPRC2012,taoPRC2013}.
Also in a relativistic microscopic approach \cite{litPRC2009} it was observed that the dipole spectra of even-even 
Ni and Sn isotopes show two well-separated collective structures, the lower one being identified with
pygmy resonance, consistent with previous results based on relativistic quasiparticle random phase approximation  (RPA)
\cite{vretenar2001,paaPRL2009, daoPRC2011}.
Other studies, however, associate the concentration of strength to the contributions from single-particle
type excitations excluding coherent, collective properties \cite{tsoPRC2008,reiPRC2013}.
It is possible that in the low-energy region the dipolar response manifests 
both single-particle and collective features. Moreover, a fragmentation of the $E1$
response is expected to determine a weakening of the collectivity \cite{sarPLB2004,gamPRC2011}.

  A promising approach aiming to clarify the nature of PDR as well as the role of the symmetry energy and the neutron
skin is based on a systematic analysis of the influence of the neutron excess on observables as the
energy centroid or the low-energy $E1$ strength.
Following this approach, several experimental investigations have been focused on the study of 
Ca \cite{harPRL2004}, Ge \cite{junNPA1995}, and Mo \cite{rusEPJA2006} isotopes as well as of 
$N=50$ \cite{schPRC2007}, and $N=82$ isotones \cite{volNPA2006}.
From the measurements for stable Sn isotopes \cite{govPRC1998,ozeNPA2007,endPRL2010,tofPRC2011} 
and neutron-rich systems $^{129,132}$Sn, and $^{133,134}$Sb 
\cite{kliPRC2007}, a trend of strength increasing
with the neutron-proton asymmetry $I^2$ was reported. A threshold value of the isospin $I$,
beyond which a sizable fraction of the pygmy strength appears, was related to the skin development
\cite{kliPRC2007}. The goal of this paper is to address the connection between the 
development of the neutron skin and the emergence of a low-energy $E1$ response in relation with the symmetry
energy density dependence, a subject under intense debate during the past few years.

Theoretically, in a semi-phenomenological description using a Hartree-Fock-Bogoliubov (HFB) treatment
within the quasiparticle-phonon model (QPM) \cite{tsoPLB2004}
for the neutron-rich Sn isotopes from $^{120}$Sn to $^{132}$Sn, it was stressed that the concentration of
$E1$ strength, evidenced between $6$ and $8 MeV$,
cannot be considered a low-energy tail of GDR. The corresponding states, having a genuine character with
a dominance of neutron excitations, were considered noncollective. 
The evolution of the strength distribution and of the energy location was 
closely related to the features of neutron mantle enclosing the more isospin symmetric core. 
However, in a non-relativistic RPA treatment \cite{coPRC2009} for zirconium isotopes, the investigation of
the role played by the neutrons in excess has shown that these strongly contribute
to the $E1$ excitation at about 8.5 MeV and make it collective. Moreover, the analysis \cite{yukNPA2012} of
neutron and proton contributions to PDR, based on a non-relativistic self-consistent HF+RPA approximation, 
indicates that the pattern of the PDR changes with the increasing neutron number, becoming a quite collective
resonant oscillation of the neutron skin. It was noticed a large collectivity of low-energy dipole states in  $^{68}$Ni
and $^{132}$Sn displaying a mixed isovector-isoscalar motion.
   
Piekarewicz \cite{piePRC2006} raised the important question if a strong
correlation between the neutron skin and the low-energy $E1$ strength 
can be distinguished. For Sn isotopes, within a relativistic RPA model, he concluded
that the fraction of EWSR acquired in the energy region between $5 MeV$ and $10 MeV$ manifests
a linear dependence with the neutron skin size  up to mass $A=120$ followed by a mild anticorrelation.
However, such strong correlation was questioned in Ref. \cite{reiPRC2010}. The authors introduced an investigation based on a 
covariance analysis aimed to identify a set of good indicators that correlate very well with the 
isovector properties and suggested that the low-energy $E1$ strength is very weakly correlated with 
the neutron skin while the dipolar polarizability should
be a much stronger indicator of isovector properties. 
This intriguing finding was challenged recently \cite{piePRC2011} in the relativistic RPA approach
with mixed results. A strong correlation between the neutron
skin thickness of $^{208}$Pb and the dipole polarizability of $^{68}$Ni was indeed reported. But a strong correlation
was also claimed between the skin thickness of $^{208}$Pb and low-energy $E1$ features, including
the strength and dipole polarizability associated to the pygmy mode, 
identified in $^{68}$Ni as exhausting about $5\%-8\%$ of the EWSR. 

Here we shall address these controversial issues, proposing an investigation based on
a semiclassical transport model. 
Because the neutron skin is an isovector 
indicator, we employ three different parametrizations with the density of the symmetry term
and perform a comparative study 
in a model based on the Landau theory of Fermi liquids where the dynamics of the nucleons is described 
by Landau-Vlasov kinetic equations.
 In this paper we first explore the properties of the neutron skin and its sensitivity to the density dependence
of symmetry energy. Then we determine the $E1$ strength function and study the mass dependence
of the PDR peak. Finally, we estimate the EWSR exhausted by
the PDR and discuss its relation with the neutron skin thickness. Since, as in the case of GDR, the evolution with mass of the 
low-energy $E1$ response provides an additional insight upon the nature of the mode, we shall 
consider the systems $^{48}$Ca, $^{68}$Ni, $^{86}$Kr and $^{208}$Pb, 
as well as a chain of Sn isotopes, $^{108,116,124,132,140}$Sn.

\section{Theoretical framework}

Having as main ingredients the fermionic nature of the constituents
and the self-consistent mean-field, the Vlasov equation represents the semiclassical limit of time-dependent Hartree-Fock (TDHF)
and, for small-oscillations, of the RPA equations.
While the model is unable to account for effects associated with the shell structure,
our self-consistent approach is suitable to describe robust quantum modes, of zero-sound type,
in both nuclear matter and finite nuclei.  It provides important information about the dynamics of such collective modes,
allowing for a systematic study over an extended mass and isospin domain.  
In this context we notice that in a TDHF study with a Skyrme interaction \cite{briIJMPE2006} a pygmy like peak was identified for the
deformed $^{34}$Mg at around 10 MeV. From time-dependent density plots it was recognized as a superimposed surface mode
not fully coupled to the bulk motion. 
Similarly, studies based on Landau-Vlasov equations were also inquiring on
the collective nature of PDR \cite{urbPRC2012,barRJP2012} 
and on the role of the symmetry energy on its
dynamics \cite{barPRC2012}. It was observed that, as in the TDHF investigation, a pygmy like collective motion in $^{132}$Sn 
manifests. Moreover, it was found that the symmetry energy does not affect the energy centroid but influences the EWSR acquired by it.

The two coupled Landau-Vlasov kinetic equations for neutrons and protons:
% $\displaystyle \frac{\partial f_q}{\partial t}+\frac{\bf p}{m}\frac{\partial f_q}{\partial {\bf r}}-
%\frac{\partial U_q}{\partial {\bf r}}\frac{\partial f_q}{\partial {\bf p}}=I_{coll}[f]$ ,
\begin{equation}
\frac{\partial f_q}{\partial t}+\frac{\bf p}{m}\frac{\partial f_q}{\partial {\bf r}}-
\frac{\partial U_q}{\partial {\bf r}}\frac{\partial f_q}{\partial {\bf p}}=I_{coll}[f_n,f_p] ,
\label{vlasov}
\end{equation}
determine the time evolution of the one-body distribution functions $f_q(\vec{r},\vec{p},t)$,
with $q=n,p$ \cite{barPR2005}. In the following we shall switch-off the collision integral but we have tested
that the results are not strongly influenced, as expected, when it is included.  
For the nuclear mean-field we consider a Skyrme-like ($SKM^*$) parametrization:
\begin{equation}
U_{q} = A\frac{\rho}{\rho_0}+B(\frac{\rho}{\rho_0})^{\alpha+1} + C(\rho)
\frac{\rho_n-\rho_p}{\rho_0}\tau_q
+\frac{1}{2} \frac{\partial C}{\partial \rho} \frac{(\rho_n-\rho_p)^2}{\rho_0}
\label{meanfield}
\end{equation}
with, $\tau_n (\tau_p)=+1 (-1)$.
The saturation properties of the symmetric nuclear matter, $\rho_0=0.16 fm^{-3}$, $E/A=-16$ MeV
and a compressibility modulus $K=200$ MeV, are reproduced if the values 
$A=-356$ MeV, $B=303$ MeV, $\alpha=1/6$ are fixed.
Concerning the density dependence of the symmetry energy, 
we consider, in the mean-field structure, different parametrizations of $C(\rho)$. While keeping the value of 
symmetry energy at saturation almost the same, we shall allow for three different dependencies with density away from equilibrium.
For the asystiff equation of state (EOS) $C(\rho)$ is constant, $ \displaystyle C(\rho) = 32$ MeV. Then the symmetry energy 
$\displaystyle E_{sym}/A = {\epsilon_F \over 3}+{C(\rho) \over 2}{\rho \over \rho_0}$ at
saturation takes the value $E_{sym}/A = 28.3$ MeV while the slope parameter 
$\displaystyle L = 3 \rho_0 \frac{d E_{sym}/A}{d \rho} |_{\rho=\rho_0}$
is $L=72$ MeV. The asysoft case corresponds to a Skyrme-like, SKM*, parametrization 
with $\displaystyle \frac{C(\rho)}{\rho_0} = (482-1638 \rho) MeV fm^3$, which leads to a small value 
of the slope parameter $L= 14.4$ MeV.
Last, for the asysuperstiff EOS, $\displaystyle \frac{C(\rho)}{\rho_0}=\frac{32}{\rho_0} \frac{2 \rho}{(\rho + \rho_0)}$, the symmetry term increases rapidly around saturation density, being characterized by a value of the slope parameter $L=96.6$ MeV.

The integration of the transport equations is based on the test-particle (or pseudoparticle) method, with a number of
$1300$ test particles per nucleon in the case of Sn isotopes, ensuring in this way a good spanning of the phase space. 
This method is able to reproduce accurately the equation of state of nuclear matter and
provide reliable results regarding the properties of nuclear surface \cite{idiNPA1993} and ground-state energy
for finite nuclei \cite{schPPNP1989}.

Since in the next section we shall explore the possible correlations between the properties of PDR and the neutron skin,
we present here the predictions of the model for the neutron and proton distributions for different asy-EOS.
From the one-body distribution functions one obtains the local densities: 
$\displaystyle \rho_q(\vec{r},t)=\int \frac{2 d^3 {\bf p}}{(2\pi\hbar)^3}f_q(\vec{r},\vec{p},t)$
%\begin{equation}
%\displaystyle \rho_q(\vec{r},t)=\int \frac{2 d^3 {\bf p}}{(2\pi\hbar)^3}f_q(\vec{r},\vec{p},t)
%\end{equation}
as well as the quadratic radii
$\displaystyle \langle r_q^2 \rangle = \frac{1}{N_q} \int r^2 \rho_q(\vec{r},t) d^3 {\bf r}$
and the width of the neutrons skin
$\displaystyle \Delta R_{np}= \sqrt{\langle r_n^2 \rangle}-\sqrt{\langle r_p^2 \rangle}=R_n-R_p$.

\begin{figure}
\begin{center}
\includegraphics*[scale=0.36]{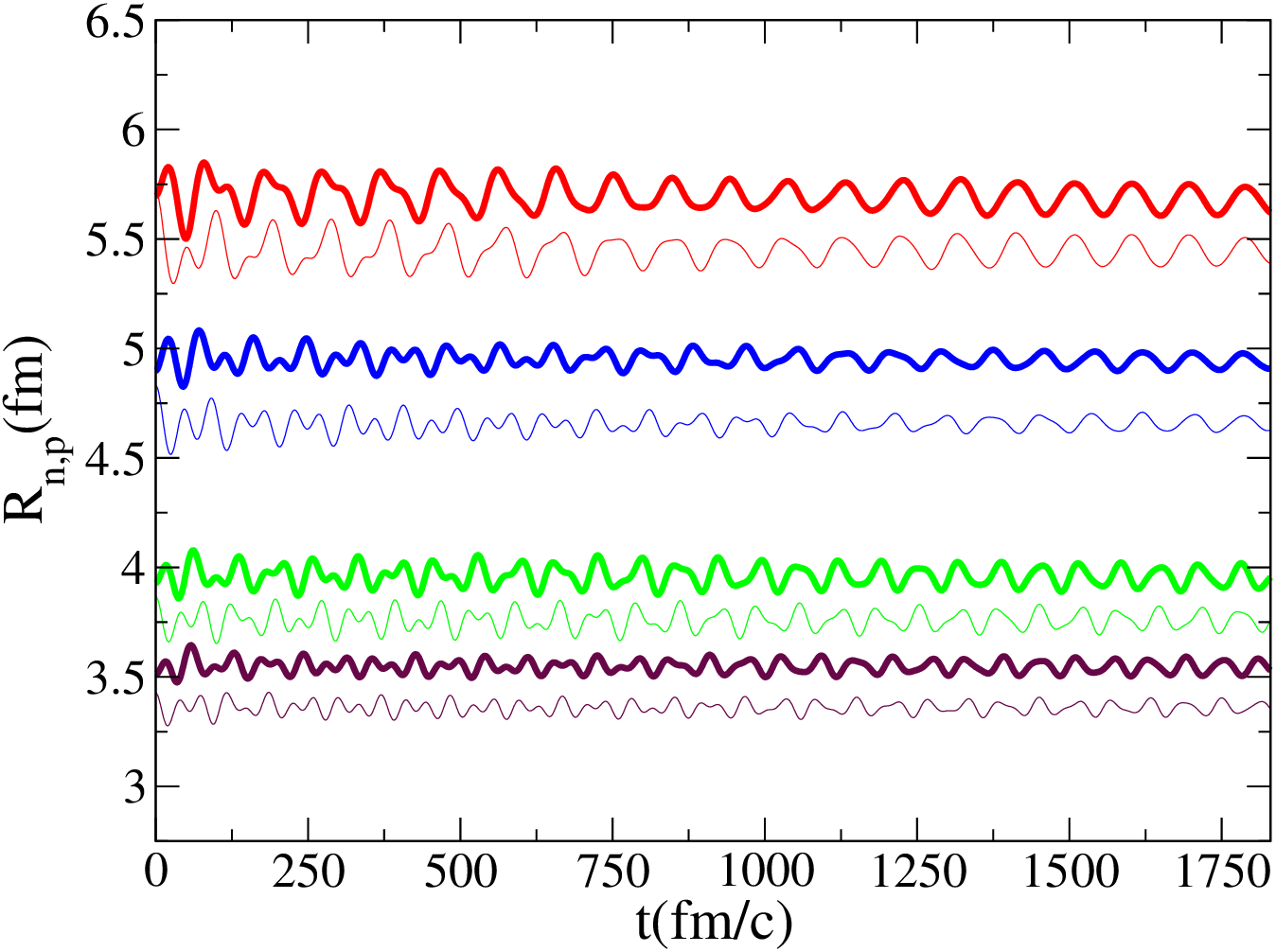}
\end{center}
\caption{The time evolution of the neutron mean square radius $R_n$ (thick lines) 
and of the proton mean-square radius $R_p$ (thin lines) after
a weak perturbation of the ground state. From the top
the pairs of lines correspond to $^{208}$Pb (red), $^{132}$Sn (blue), $^{68}$Ni (green)
and $^{48}$Ca (maroon). The asystiff EOS case.}
\label{rad_time}  
\end{figure}

\begin{figure}
\begin{center}
\includegraphics*[scale=0.46]{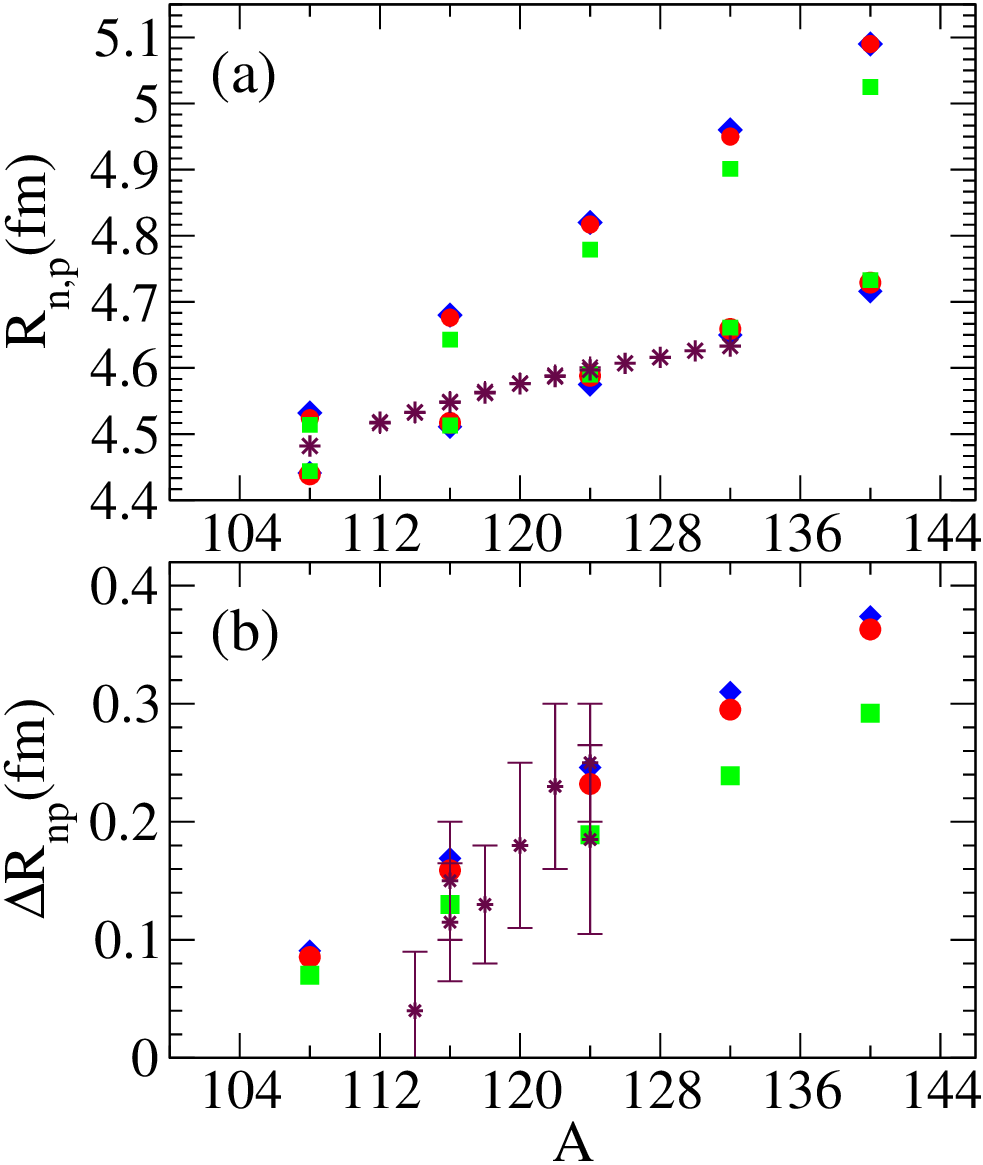}
\end{center}
\caption{(Color online) (a) The neutron and proton mean-square radius
for Sn isotopes: asysoft (the green squares), asystiff (the red circles), and asysuperstiff (the blue diamonds) EOS.
The stars are experimental data from Refs. \cite{jagADN1987,angADN2013}.
(b) The neutron skin thickness as a function of mass for Sn isotopes: asysoft (green squares),
asystiff (the red circles), and asysuperstiff (the blue diamonds). The stars and the error bars (maroons)
are experimental data from Ref. \cite{kraPRL1999}. }
\label{radius}
\end{figure}

An efficient method to extract the values of  $R_n$ and $R_p$
is by observing their time evolution after a gentle monopolar perturbation.
Both quantities perform small oscillations
around equilibrium values and we remark that the numerical simulations keep
a very good stability of the dynamics for at least $1800$ fm$/c$, see Fig. \ref{rad_time}. 
Using this procedure, we obtain for the charge mean square radius of $^{208}Pb$ a value around $R_p=5.45$ fm, to be compared with
the experimental value $R_{p, exp}=5.50$ fm. For Sn isotopes we display the mass dependence of 
$\displaystyle R_n, R_p$ in Fig. \ref{radius}-(a) and of $\displaystyle \Delta R_{np}$, respectively, 
in Fig. \ref{radius}-(b). The charge radii predictions from the three asy-EOS virtually
coincide and we notice a good agreement with the experimental data reported in Refs. \cite{jagADN1987,angADN2013}.
However, the calculations somehow underestimate the charge radii 
at smaller $A$ and tend to overestimate it towards larger A, thus providing a stronger rise tendency than observed experimentally.
For all adopted parametrizations the values of the neutron skin thickness
are within the experimental errors bars, see the data presented in Ref. \cite{kraPRL1999} for the stable Sn nuclei.
In the case of $^{208}$Pb we find $\Delta R_{np}=0.19$ fm for asysoft, 
$\Delta R_{np}=0.25$ fm for asystiff, and $\Delta R_{np}=0.27$ fm for asysuperstiff EOS while for 
$^{68}$Ni the corresponding values are $\Delta R_{np}=0.17, 0.19, 0.20$ fm. As expected, the neutron skin thickness increases
with the slope parameter $L$, an effect related to the tendency of the system to stay more isospin symmetric even 
at lower densities when the symmetry energy changes slowly below saturation, as in the case of the asysoft EOS.

\section{Collective Pygmy Dipolar Response}
\begin{figure}
\begin{center}
\includegraphics*[scale=0.46]{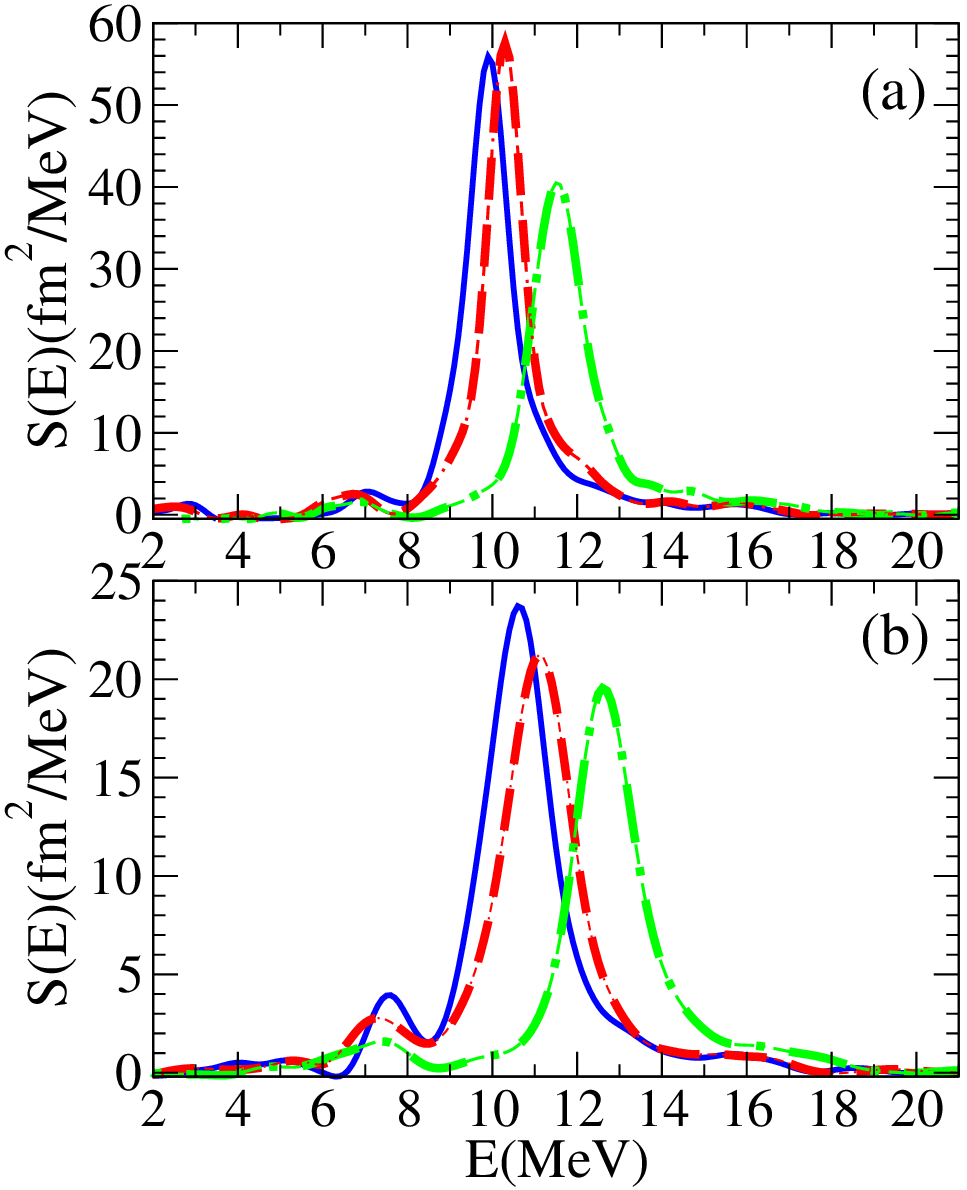}
\end{center}
\caption{(Color online) The strength function for $^{208}$Pb (a) and $^{140}$Sn (b)
for asysoft [the green (dot-dashed) lines], aystiff [the red (dashed) lines], and asysuperstiff [the blue (solid) lines]
EOS.}
\label{strengths}
\end{figure}

We study the $E1$ response considering a GDR-like initial
condition \cite{barPRC2012}, determined by the instantaneous
excitation $\displaystyle V_{ext} =\eta \delta(t-t_0) \hat{D}$ at $t=t_0$ \cite{calAP1997}. 
This situation corresponds to a boost of all neutrons against all protons 
while keeping the center of mass (c.m.) at rest.  
Here $\hat{D}$ is the dipole operator. If $|\Phi_{0} \rangle$ is the
state before perturbation then the excited state becomes $\displaystyle |\Phi (t_0)\rangle 
=e^{i \eta \hat{D}} |\Phi_{0} \rangle$. The value of $\eta$ can be related to the initial 
expectation value of the collective dipole momentum $\hat{\Pi}$,
\begin{equation}
\langle \Phi (t_0)|\hat{\Pi}|\Phi (t_0) \rangle  = \hbar \eta \frac{N Z}{A}.
\label{eta}
\end{equation}
Here $\hat{\Pi}$ is canonically conjugated to the collective coordinate $\hat{X}$, which defines the 
distance between the center of mass of protons and the center of mass of neutrons, 
i.e., $[\hat{X},\hat{\Pi}]=i\hbar$ \cite{barRJP2012}.
Then the strength function
\begin{equation}
S(E)=\sum_{n > 0}|\langle n|\hat{D}|0\rangle|^2\delta(E-(E_n-E_0)) ,
\end{equation} 
where $E_n$ are the excitation energies of the states $|n\rangle$ while
$E_0$ is the energy of the ground state $\displaystyle |0\rangle=|\Phi_{0} \rangle$, is obtained  
in our approach from the imaginary part of the Fourier transform of the time-dependent expectation value of 
the dipole momentum $ \displaystyle D(t) = \frac{NZ}{A} X(t)= \langle \Phi (t) |\hat{D}| \Phi (t) \rangle $ as:
%$ \displaystyle S(E) =\frac{Im(D(\omega))}{\pi \eta \hbar}$,
\begin{equation}
 S(E) =\frac{Im(D(\omega))}{\pi \eta \hbar}~~,
\label{stre}
\end{equation}
where $\displaystyle D(\omega) =\int_{t_0}^{t_{max}} D(t) e^{i\omega t} dt$.
We consider the initial perturbation along the $z$ axis and follow the dynamics of the system until $t_{max}=1830$ fm$/c$.
At $t=t_0=30$ fm$/c$ we extract the collective momentum and determine $\eta$. 
A filtering procedure, as described in Ref. \cite{reiPRE2006}, was applied in order to eliminate the artifacts resulting from a finite time domain analysis of the signal. A smooth cut-off function was introduced such
that $D(t) \rightarrow D(t)cos^{2}(\frac{\pi t}{2 t_{max}}) $. 
For the three asy-EOS the $E1$ strength functions of $^{208}$Pb and $^{140}$Sn 
are represented in Fig. \ref{strengths}. As a test of the quality of our method we compared the numerically estimated
value of the first moment $\displaystyle m_1=\int_0^\infty E S(E) dE$  with the value predicted by the
Thomas-Reiche-Kuhn (TRK) sum rule $\displaystyle m_1= \frac{\hbar^2}{2m} \frac{N Z}{A}$. In all cases 
the difference was only a few percentages.

\begin{figure}
\begin{center}
\includegraphics*[scale=0.46]{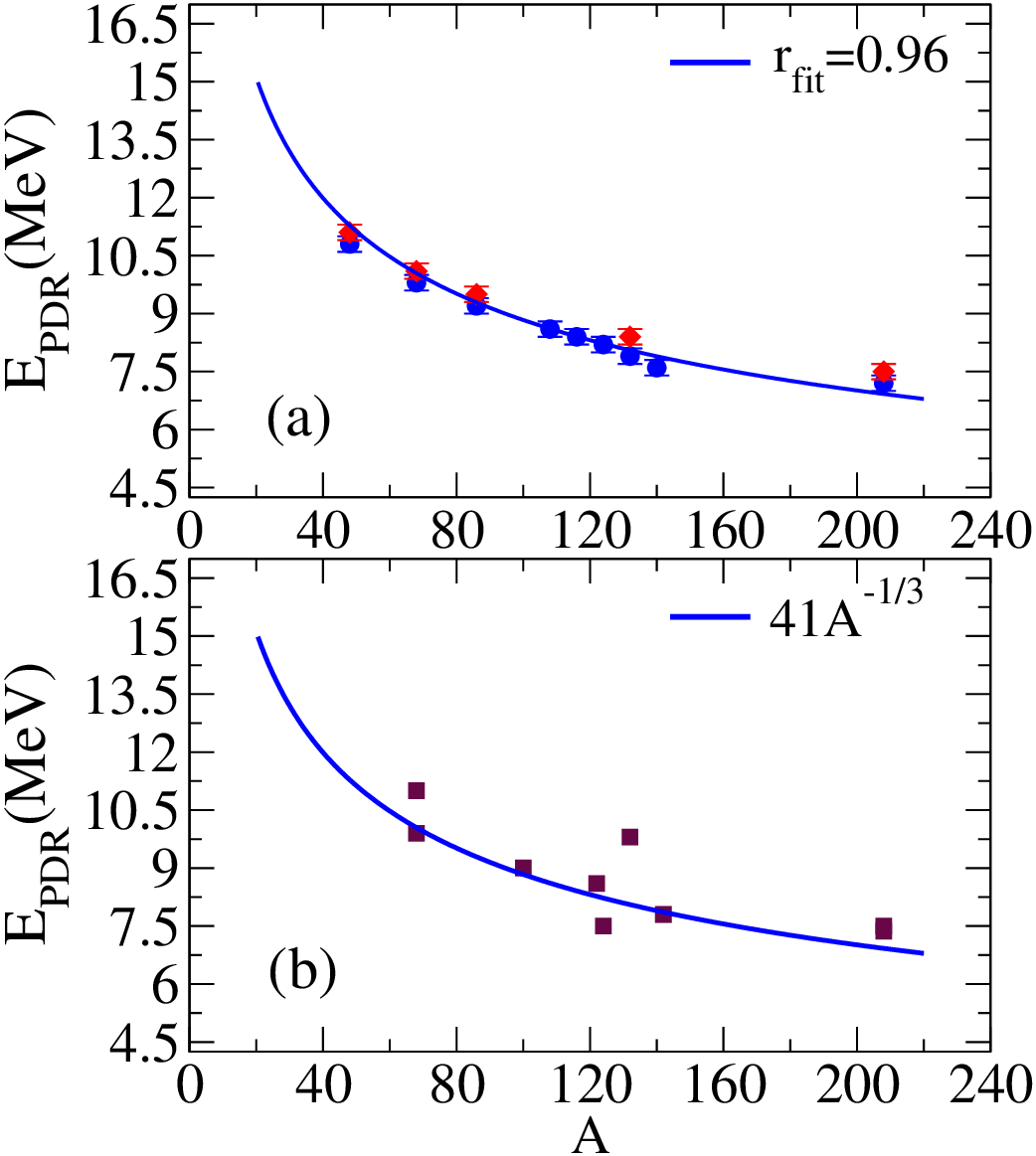}
\end{center}
\caption{(Color online) (a) The energy centroid of PDR as a function of mass. The blue circles and red
diamonds are the predictions of the model; see the text. 
The best fit, the solid (blue) line, corresponds to the parametrization $E_{PDR}=41 A^{-1/3}$.
$r_{fit}$ refers to the correlation coefficient.
(b) The energy centroid of PDR from experimental data. 
The maroon squares are experimental data points, see the text.
The solid blue line corresponds to the parametrization $41 A^{-1/3}$.
}
\label{centroid}
\end{figure}

The energy peak of the PDR for $^{208}$Pb, see Fig. \ref{strengths}(a), is 
located around $7-7.5$ MeV  in good agreement with experimental data which indicate
$E_{PDR,Pb}=7.36$ MeV \cite{tamPRL2011}. For $^{68}$Ni we obtain $9.8$ MeV, quite close to
the recent reported data $E_{PDR,Ni}=9.9$ MeV  \cite{rossi2013}.
We observe that the GDR energy centroid is underestimated in comparison with experimental data, a fact related with the choice
of the interaction which has not an effective mass 
\cite{surNPA1988}. In any case, a clear dependence on the slope parameter manifests as a consequence 
of the isovector nature of the mode.
This feature shows that also the symmetry energy values below saturation
are affecting the dipole oscillations of the finite systems.
Figure \ref{centroid}(a) displays the predicted
position of the PDR energy centroid as a function of mass for all studied systems (blue circles).
Since in all cases a very weak influence of the symmetry energy on the PDR peak was observed, we take the average
of the values corresponding to the three asy-EOS. Then the error bars represent the deviation of the
determined values from the average.
In addition, by using the same procedure, we represent the position of the PDR energy peaks as results from the power
spectrum analysis of the pygmy dipole $D_y(t)$ after a pygmy like initial condition, see Ref. \cite{barPRC2012} 
(red diamonds). The differences between the two methods are within a half of a MeV. 
An appropriate parametrization, obtained from the fit of numerical results 
is
\begin{equation}
E_{PDR}=41A^{-1/3} MeV ~,
\end{equation}
quite close to what is expected in the harmonic
oscillator shell model (HOSM) approach \cite{barRJP2012}. In Fig. \ref{centroid}(b) this 
parametrization is compared with the experimental
data available from the works where information about the position of the low-energy $E1$ centroid
was reported (maroon square) 
\footnote{For $^{68}$Ni from Refs. \cite{wiePRL2009}, \cite{rossi2013}; for $^{100}$Mo from Ref. \cite{rusEPJA2006};
for $^{122}$Sn from Ref. \cite{tofPRC2011}; for $^{124}$Sn from Ref. \cite{endPRL2010}; for $^{132}$Sn from Ref. \cite{adrPRL2005};
for $^{142}$Nd from Ref. \cite{angPRC2012}; for $^{208}$Pb from Refs. \cite{tamPRL2011,konPRC2012}.}.  
The formula seems to describe quite well the position 
of the low-energy centroid observed experimentally for several
systems. While the isovector residual 
interaction pushes up the value of the GDR energy, it seems that the PDR energy centroid is not much
affected by this part of the interaction. This feature may explain the better agreement with experimental
observations in comparison with the GDR case. The agreement also suggests that the PDR peak energy 
should not be significantly influenced by the momentum dependence of the interaction. 
Let us mention that for Ni, Sn, and Pb isotopic chains, based on a HFB and RQRPA treatment,
Paar \emph{et al.} \cite{paarPLB2005} studied the isotopic dependence of the PDR energy and   
a collective mode with the energy centroid around $10$ MeV for $^{68}$Ni, $8$ MeV for $^{132}$Sn and
$7.5$ MeV for $^{208}$Pb was predicted. A comparison with our results
shows a good concordance between the two theoretical approaches for the position of the PDR energy centroid.

\begin{figure}
\begin{center}
\includegraphics*[scale=0.36]{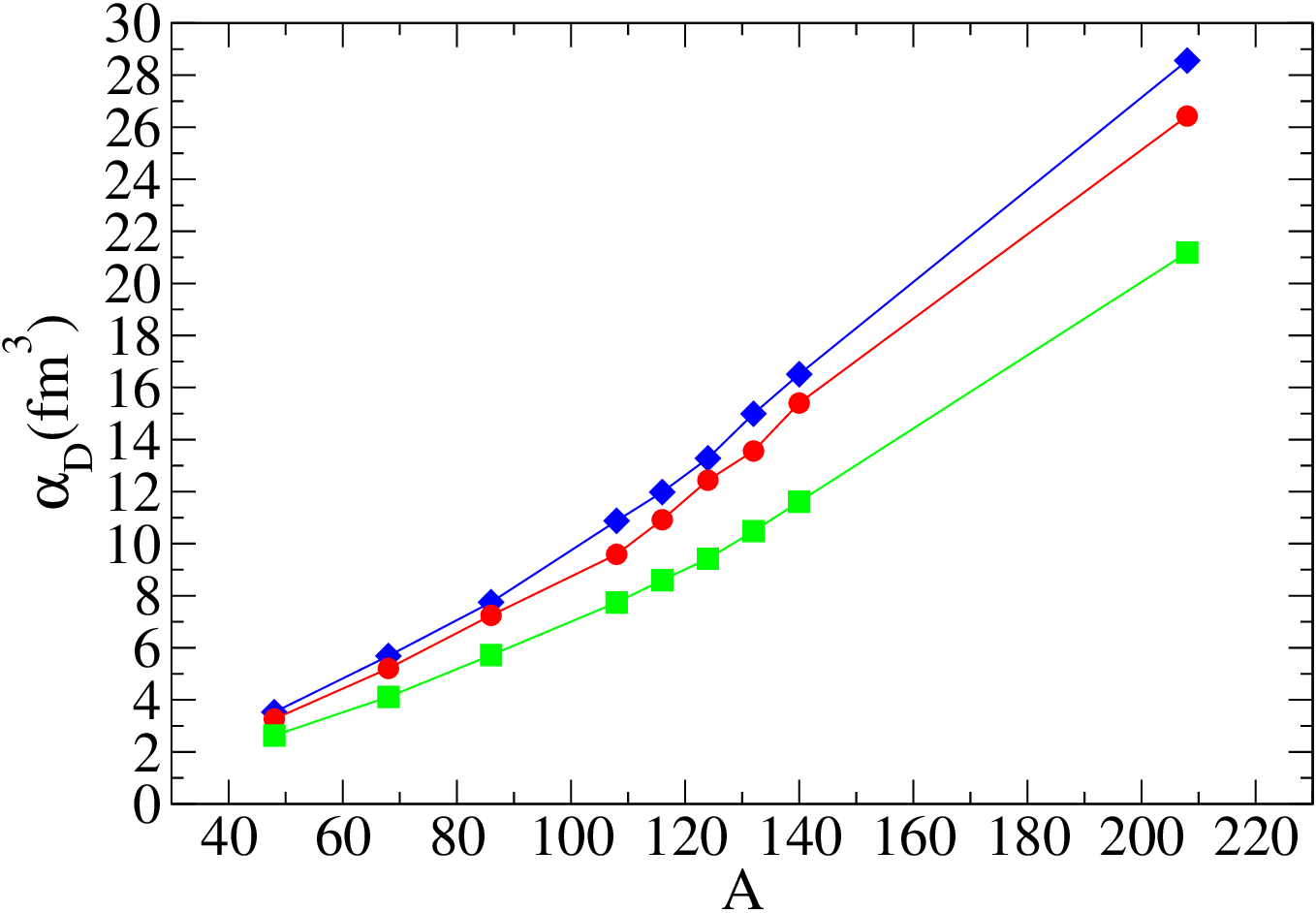}
\end{center}
\caption{(Color online) The dipole polarizability as a function of mass 
for asysoft (the green squares), asystiff (the red circles), and asysuperstiff (the blue diamonds) EOS.
All systems mentioned in text are included.}
\label{polariz}
\end{figure}

Having obtained the strength function, we can calculate the nuclear dipole polarizability,
%$ \displaystyle \alpha_D = 2 e^2 \int_0^{\infty}\frac{S(E)}{E} dE$ 
\begin{align}
\alpha_D = 2 e^2 \int_0^{\infty}\frac{S(E)}{E} dE  ~,
\end{align}
as an additional test of the numerical method. In the case of $^{68}$Ni  $\displaystyle \alpha_D$ 
varies from  $4.1$ fm$^3$ to $5.7$ fm$^3$ when we pass from asysoft to asysuperstiff EOS
while for $^{208}$Pb it changes from $21.1$ fm$^3$ to $28.6$ fm$^3$. Since for proposed interactions
the position of the GDR peak is below the experimental observations, we expect that the values of the polarizabilities
to be somehow overestimated in comparison with data. Experimentally, the dipole polarizability is 
below $4$ fm$^3$ for $^{68}$Ni \cite{rossi2013} 
and around $20$ fm$^3$ for $^{208}$Pb \cite{tamPRL2011}. We display this quantity as a 
function of mass and asy-EOS in Fig. \ref{polariz}. 
For a given system, the larger the neutron skin thickness,
the greater the value of the dipole polarizability obtained.

\begin{figure}
\begin{center}
\includegraphics*[scale=0.36]{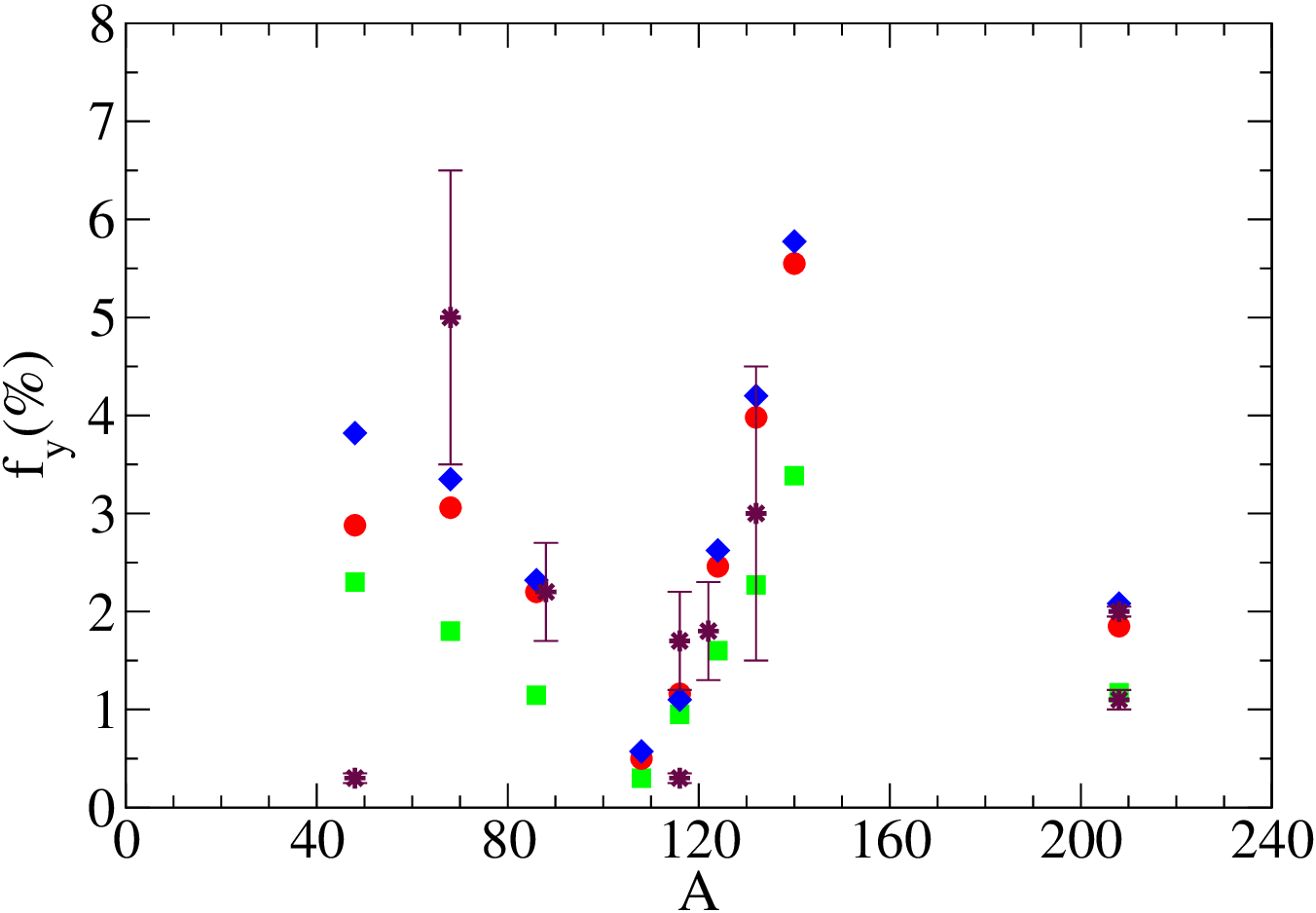}
\end{center}
\caption{(Color online) The fraction of EWSR exhausted by PDR as a function of mass for asysoft (green squares),
asystiff (red circles), and asysuperstiff (blue diamonds) EOS for the systems $^{48}$Ca, $^{68}$Ni, $^{86}$Kr,
$^{108}$Sn, $^{116}$Sn, $^{124}$Sn, $^{132}$Sn, $^{140}$Sn, $^{208}$Pb. 
The stars (maroon) are experimental data points obtained by various methods for $^{48}$Ca, $^{68}$Ni, $^{88}$Sr, $^{116}$Sn,
$^{122}$Sn,$^{132}$Sn, $^{208}$Pb, reported in Ref.\cite{savPPNP2013}.}
\label{fy}
\end{figure}

The EWSR exhausted by the PDR is calculated by integrating over the low-energy resonance region as follows:
\begin{equation}
 m_{1,y} = \int_{PDR} E S(E) dE  ~.
\end{equation}
We carefully determined the limits of the pygmy resonance region, identifying the minima of the response around the PDR centroid.
When an overlap with the GDR region exists, the contribution from the GDR tail is substracted.
In Fig. \ref{fy} the fraction $\displaystyle f_y=\frac{m_{1y}}{m_1}$ exhausted by the pygmy dipole resonance
as a function of mass is reported. Some comments are valuable when we compare our results with experimental data 
concerning the same quantity obtained from various experiments,  as presented in Fig. 26 of Ref. \cite{savPPNP2013}.
For $^{48}$Ca the fraction is obtained experimentally from the strength observed to $10$ MeV and is below $0.3\%$. However,
our calculations point out that the PDR is mainly above $10$ MeV and we obtain a fraction, depending on the asy-EOS,
between $2.3\%$ and $3.8\%$. For $^{68}$Ni, our model, with a calculated fraction
between $1.8\%$ and $3.5\%$, underestimates the experimental value, which is around $f_y=5\%$. 
In the mass region $A=88-90$ the experimental fraction is situated at $f_y=2\%$ and for $^{86}$Kr we obtain
between $1.1\%$ and $2.3\%$. For the stable Sn isotopes, in the mass region $A=116-124$, the fraction measured experimentally
is between $1.2\%$ and $2.2\%$ while our calculations suggest values between $0.95\%$ and $1.1\%$ for  $^{116}$Sn and between
$1.6\%$ and $2.6\%$ for $^{124}$Sn. In the case of $^{132}$Sn our results are between $2.2\%$ and $4.2\%$ while experimentally
the reported fraction is around $5\%$. Finally, for $^{208}$Pb we obtain a fraction between $1\%$ and $2\%$, which is the same range
as that in the existing experimental data. We conclude that, despite the fact that our approach
has the tendency to underestimate the experimental results (except for Calcium, for the reasons discussed above),
the model reveals that a substantial part of the total fraction $f_y$ exhausted in the low-energy region 
can be attributed to the collective PDR.   

\begin{figure}
\begin{center}
\includegraphics*[scale=0.36]{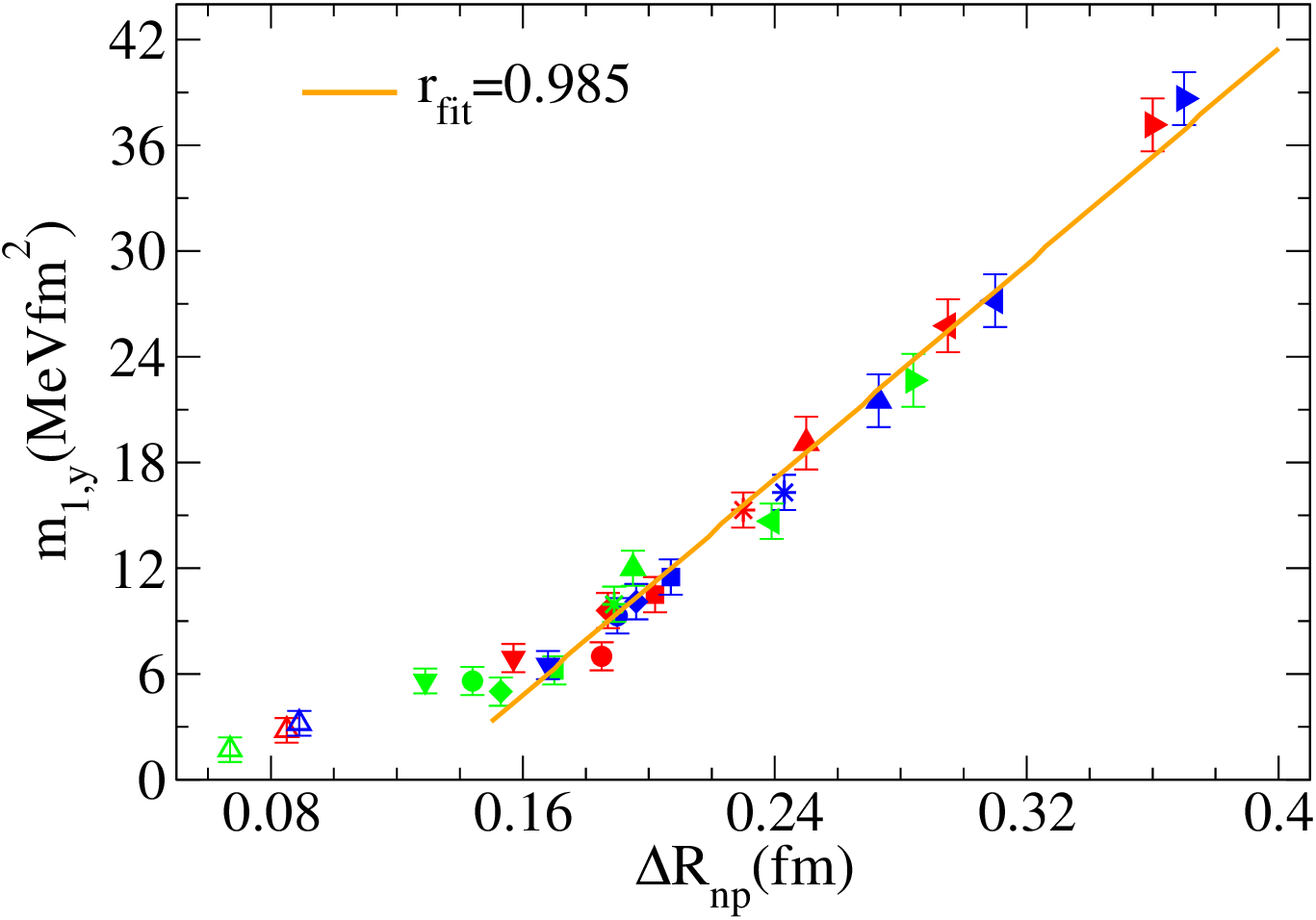}
\end{center}
\caption{(Color online) The EWSR exhausted by PDR as a function of neutron skin for
$^{108}$Sn (empty up-triangles), $^{116}$Sn (down-triangles), $^{124}$Sn (stars),
$^{132}$Sn (left triangles), $^{140}$Sn (right triangles), $^{48}$Ca (circles),
$^{68}$Ni (squares), $^{86}$Kr (diamonds), $^{208}$Pb (full up- triangles)
for asysoft (green), asystiff (red), and asysuperstiff (blue) EOS. The error bars are 
related to the uncertainties in defining the integration domain for the PDR response.
$r_{fit}$ refers to the correlation coefficient.}
\label{m1y}
\end{figure}

We investigate now if some correlation between the absolute value of EWSR exhausted in the PDR region and the
development of neutron skin manifests in our approach.
The dependence of the moment $\displaystyle m_{1,y}$ on the neutron skin thickness is shown in Fig. \ref{m1y},
where the information concerning  all mentioned systems was included. The error bars are associated 
with the uncertainty in the identification of the limits of the pygmy resonance region.
%{\bf In each case we designate a minimal and a maximal %extension of the PRD region 
%which determine the reported error bars.} 
While below $0.15 fm$ the EWSR acquired by the PDR manifests a saturation tendency, above this value
a linear correlation arises. For the same nucleus, when we pass from the asysuperstiff
to asysoft parametrization, the neutron-skin shrinks and, correspondingly, the value of 
$m_{1,y}$ decreases. This behavior is in agreement with the results reported in Ref. \cite{vrePRC2012}
in a self-consistent RPA approximation based on relativistic energy density functionals.
Moreover, we notice that the variation rate appears to be system independent, obtaining an increase
of $15$ MeV fm$^2$ of the exhausted EWSR, versus a change of $0.1 fm$ of the neutron skin width.
Such features suggest that the acquired EWSR should not differ too much even for different nuclei 
if they have close values of neutrons skin thickness. These findings look 
qualitatively in agreement with those of Inakura \emph{et al.}  
\cite{inaPRC2011}, based on systematic calculations within a
RPA treatment with a SkM* Skyrme functional, where a linear correlation of the fraction of EWSR exhausted in the 
the energy region up to $10$ MeV and neutron skin thickness was evidenced for several isotopic chains. 

%\begin{figure}
%\begin{center}
%\includegraphics*[scale=0.26]{m1y_rnp_lett.eps}
%\end{center}
%\caption{(Color online) The EWSR exhausted by PDR as a function of neutron skin in
%$^{108}Sn$ (empty triangles up), $^{116}Sn$ (triangles down), $^{124}Sn$ (stars),
%$^{132}Sn$ (triangles left), $^{140}Sn$ (triangles right), $^{48}Ca$ (circles),
%$^{68}Ni$ (squares), $^{86}Kr$ (diamonds), $^{208}Pb$ (full triangles up)
%for asysoft (green), asystiff (red) and asysuperstiff (blue) EOS. $r_{fit}$ refers
%to the correlation coefficient. The error bars are 
%related to the uncertainties in defining the integration domain for the PDR response.}
%\label{ewsr}
%\end{figure}

However, some differences are also worth mentioning. While we observe the total amount of EWSR
exhausted in the PDR region, $m_{1y}$,  manifests a system independent, linear dependence 
with the neutron skin thickness, with a slope $s=150 MeVfm$,
Inakura \emph{et al.} deduce a linear correlation of the fraction $f_y$ as a function of $\Delta R_{np}$. In this case, the corresponding 
universal rate is  $0.18-0.20$ fm$^{-1}$ for even-even nuclei with $8\le Z\le 40$. To establish a connection between the two
approaches, we shall assume that within a specific isotopic chain the ratio $NZ/A$ does not change too much, i.e., the
value of $m_1$ is approximately the same for all those nuclei. With this approximation, for a fixed isotopic chain, the 
two predictions are similar, i.e. an universal slope for $f_y$ is equivalent with an universal slope
for $m_{1y}$. Consequently, in the Inakura approach, it can be deduced that the value of the slope $s$ is 
around $70$ MeV fm for Ca isotopes, $95$ MeV fm for Ni chain, and $120$ MeV fm for Kr isotopes. 

We also remark that, in Ref. \cite{inaPRC2011}, for very neutron rich systems 
a mild anti-correlation of $f_y$ with the neutron skins begin to manifest, similarly to 
the results of Refs. \cite{piePRC2006,liaPRC2007}.
This feature is missing in our model. We obtain a continuous rise of $m_{1,y}$
with the neutron skin size, in concordance with other studies based on microscopic treatments 
\cite{tsoPLB2004,penPRC2009}.
One can relate these differences to some shell and angular momentum effects but further 
investigations are required for a definite answer.

\section{Conclusions}

Summarizing, we addressed some of the open questions raised recently \cite{savPPNP2013} regarding the nature of the PDR.
By performing a systematic investigation over an extended mass domain,
new features, providing a more complete picture of the PDR dynamics, were evidenced.
In a microscopic transport approach, a low-energy dipole collective mode occurs as an ubiquitous property 
of all investigated systems. The analysis leads us
to a dependence of the PRD energy centroid with mass very well described 
by the parametrization $E_{PDR}=41 A^{-1/3}$,
in agreement with several recent experimental results. This indicates a close connection with
the distance between major shells, $\hbar \omega_0 =41 A^{-1/3}$, and 
a weak influence of the residual interaction in the isovector sector.
Such behavior can be related to  the isoscalar-like nature of this mode. 
We notice that the EWSR exhausted by the collective pygmy dipole depends
on the symmetry energy slope parameter $L$ and
represents a significant part of the value determined experimentally.
From our investigation, an universal, linear correlation
of $m_{1y}$ with the neutron skin thickness emerges. 
It appears as a very specific signature, showing that 
the neutrons which belong to the skin play an essential
role in shaping the $E1$ response in the PDR region. 
However, this fact should not lead to an oversimplified picture
of the PDR, as corresponding only to the oscillations of the excess neutrons against an inert isospin symmetric core. Within our transport model,  
the dynamical simulations show a more complex structure 
of the PDR \cite{barPRC2012},
which includes an isovector excitation of the core and the neutrons skin oscillation.
We consider that the new findings presented here can be useful for
further, systematic experiments searching for this, quite elusive, mode.
A precise estimate of the EWSR acquired by the PDR can provide indications about the neutron skin 
size, which in turn will add more constraints on the slope parameter $L$ of the symmetry energy.

\section{Acknowledgments}
V.B. and A.C. were supported by a grant from the Romanian National
Authority for Scientific Research, CNCS - UEFISCDI, Project No. PN-II-ID-PCE-2011-3-0972.

\end{document}